\documentclass[a4paper]{elsarticle}

\usepackage{graphics}
\usepackage{amssymb}
\usepackage{color} %% pour corrections
\usepackage{pstricks}
\usepackage[ULforem]{ulem}

\journal{Computers \& Fluids}

\begin{document}

\begin{frontmatter}



\title{A reference solution of the flow over a circular cylinder at $Re=40$}


\author{R\'emi Gautier}
\author{Damien.Biau}
\author{Eric Lamballais\corref{cor1}}
\ead{eric.lamballais@univ-poitiers.fr}
\address{Institut Pprime,\\
CNRS-Universit\'e de Poitiers-ENSMA,\\
T\'el\'eport 2-Bd. Marie et Pierre Curie B.P. 30179\\
86962 Futuroscope Chasseneuil Cedex, France}
\cortext[cor1]{Corresponding author}

\begin{abstract}
%New perspectives in the last decades are offered with immersed boundary methods for flow simulations over solid body. Until now, the best numerical accuracy obtained is order two in space, independently of the numerical scheme used. In order to distinguish in the various theoretical improvements we felt necessary to propose a reference solution. The academic steady flow around a circular cylinder at Reynolds number 40 has been retained. 
The classical problem of the flow over a circular cylinder at Reynolds number 40 is considered using an accurate pseudo-spectral code. A new set of boundary conditions is proposed to improve the representation of the infinite flow domain, especially in the far wake area. It is shown that the resulting accuracy of the computed flow allows its use as a reference solution for code validation. This reference solution is reachable at any location up to 50 cylinder diameters far from the cylinder centre through spectral interpolation with a user-friendly script provided in appendix. It is shown how this solution offers the opportunity to perform a convergence study and to investigate the spatial distribution of numerical errors. The main goal of this study is to propose this reference solution as an helpful tool for numerical validation and development, especially for the improvement of immersed boundary methods toward high-order accuracy.
\end{abstract}


\end{frontmatter}

% \linenumbers

\section{Introduction\label{introduction}}

In the context of Computational Fluid Dynamics (CFD), the certification of numerical methods is often challenging. A major difficulty for validation is connected to the lack of exact solution that can be used as reference. More precisely, among the known analytical solutions of Navier-Stokes equations, no one can represent faithfully complex phenomena involving for instance vortex dynamics or flow separation. However, these phenomena play a major role in real-life flows, so that the ability of a given numerical code to predict them is a crucial issue. 

To increase the validation framework, exact solutions can be manufactured, but again, the resulting analytical solution is never fully representative of a complex flow situation. To address realistic flow configuration, validation is commonly performed by comparison with experimental or numerical results of reference. This method can be seen as the ultimate step of code verification in the sense that the overall influence of numerical parameters can be assessed practically. However, this validation strategy suffers from important drawbacks. First, because the error is global, it is difficult to distinguish each type of error contribution, for example those associated with differentiation from those connected to the boundary condition approximations. Secondly, comparisons cannot be very accurate in the sense that they are based on a limited number of reference quantities submitted to uncertainty. This uncertainty can be due to an incomplete knowledge of boundary conditions (experiments and calculations) or to measurement errors (experiments). It prevents the rigorous estimation of the convergence rate toward a solution assumed free from errors. Even the simple comparison between results from different numerical codes can be ambiguous because any conclusion has to be drawn modulo a significant range of uncertainty.

To illustrate this difficulty, let us consider one of the most popular test cases in CFD: the flow over a circular cylinder at $Re=40$. Here, $Re=UD/\nu$ is the Reynolds number based on the upstream constant velocity $U$ and the cylinder diameter $D$ while $\nu$ is the kinematic viscosity of the fluid. For this low Reynolds number, the flow is two-dimensional (2D), symmetric and steady. 
These three properties should allow straightforward and computationally inexpensive comparisons. However, the resulting flow pattern is not trivial due to the presence of two separations leading to the formation of two recirculation bubbles behind the cylinder in the near-wake region. The right prediction of the separation location is a key component to get an accurate numerical solution because no singularity in the geometry can fix the separation start, as it would be for instance for a square cylinder.

In a first view, the flow over a cylinder at $Re=40$ seems to be a well defined and discriminatory test flow for CFD. This is probably the reasons that led many previous authors to consider this flow configuration for validation. However, in the literature, a significant scattering of results can be observed, even for global quantities like the drag coefficient $C_D$ or the recirculation length $L_w$. For instance, if we consider the set of reference results \cite{tritton59,dennis&chang70,coutanceau&bouard77,fornberg80,he&doolen97,yeetal99,calhoun02,russel&wang03,tseng&ferziger03b,linnick&fasel05,chung06,leetal06,dingetal07,taira&colonius07,Posdziec,patil&lakshisha09,bouchonetal12}, we observe $2.13<L_w<2.35$ and $1.48<C_D<1.62$ (see table \ref{tab_2}). The corresponding relative uncertainty of about 10\% forbids any accurate validation simply based on these values. In addition to measurement errors for experiments or numerical errors for simulations, a fundamental reason of this scattering is connected to the challenge of reproducing, numerically or experimentally, the flow in an unbounded domain. Another explanation of the experiment/simulation discrepancy could also be due to the difficulty to establish experimentally a purely 2D flow.

\begin{table}
\caption{Physical parameters of the flow pattern around a circular cylinder at $Re=40$: Drag coefficient $C_D$, 
separation angle $\theta_s$, wake length $L_w/D$ and location of recirculation centre $(a,b)$.}
\label{tab_2}
\begin{center}
\begin{tabular}{rccccc}
\hline
                                                  & $C_D$ & $\theta_s$    & $L_w/D$ & $a/D$ & $b/D$ \\
\hline
Tritton {\cite{tritton59}}                        & 1.48   &               &         &       &       \\
Dennis \& Chang {\cite{dennis&chang70}}           & 1.52   & 126.2$^\circ$ & 2.35    &       &       \\ 
Coutanceau \& Bouard {\cite{coutanceau&bouard77}} &        & 126.2$^\circ$ & 2.13    & 0.76  & 0.59  \\ 
Fornberg {\cite{fornberg80}}                      & 1.50   & 124.4$^\circ$ & 2.24    &       &       \\
He \& Doolen {\cite{he&doolen97}}                 & 1.50   & 127.2$^\circ$ & 2.25    &       &       \\
Ye et al. {\cite{yeetal99}}                       & 1.52   &               & 2.27    &       &       \\
Calhoun {\cite{calhoun02}}                        & 1.62   & 125.8$^\circ$ & 2.18    &       &       \\
Russel \& Wang {\cite{russel&wang03}}             & 1.60   &               & 2.29    &       &       \\
Tseng \& Ferziger {\cite{tseng&ferziger03b}}      & 1.53   &               & 2.21    &       &       \\
Linnick \& Fasel {\cite{linnick&fasel05}}         & 1.54   & 126.4$^\circ$ & 2.28    & 0.72  & 0.60  \\
Chung {\cite{chung06}}                            & 1.54   &               & 2.30    &       &       \\
Le et al. {\cite{leetal06}}                       & 1.56   &               & 2.22    &       &       \\
Ding et al. {\cite{dingetal07}}                   & 1.58   & 127.2$^\circ$ & 2.35    &       &       \\
Taira \& Colonius {\cite{taira&colonius07}}       & 1.54   & 126.3$^\circ$ & 2.30    & 0.73  & 0.60  \\       
Posdziec \& R. Grundmann {\cite{Posdziec}}        & 1.49   &               &         &       &       \\
Patil \& Lakshisha {\cite{patil&lakshisha09}}     & 1.56   & 127.3$^\circ$ & 2.14    &       &       \\
Bouchon et al. {\cite{bouchonetal12}}             & 1.50   & 126.6$^\circ$ & 2.26    & 0.71  & 0.60  \\
Present reference solution                        & 1.49   & 126.4$^\circ$ & 2.24    & 0.71  & 0.59  \\
\hline
\end{tabular}
\includegraphics[bb= 0 170 842 480,width=0.9\textwidth,clip=]{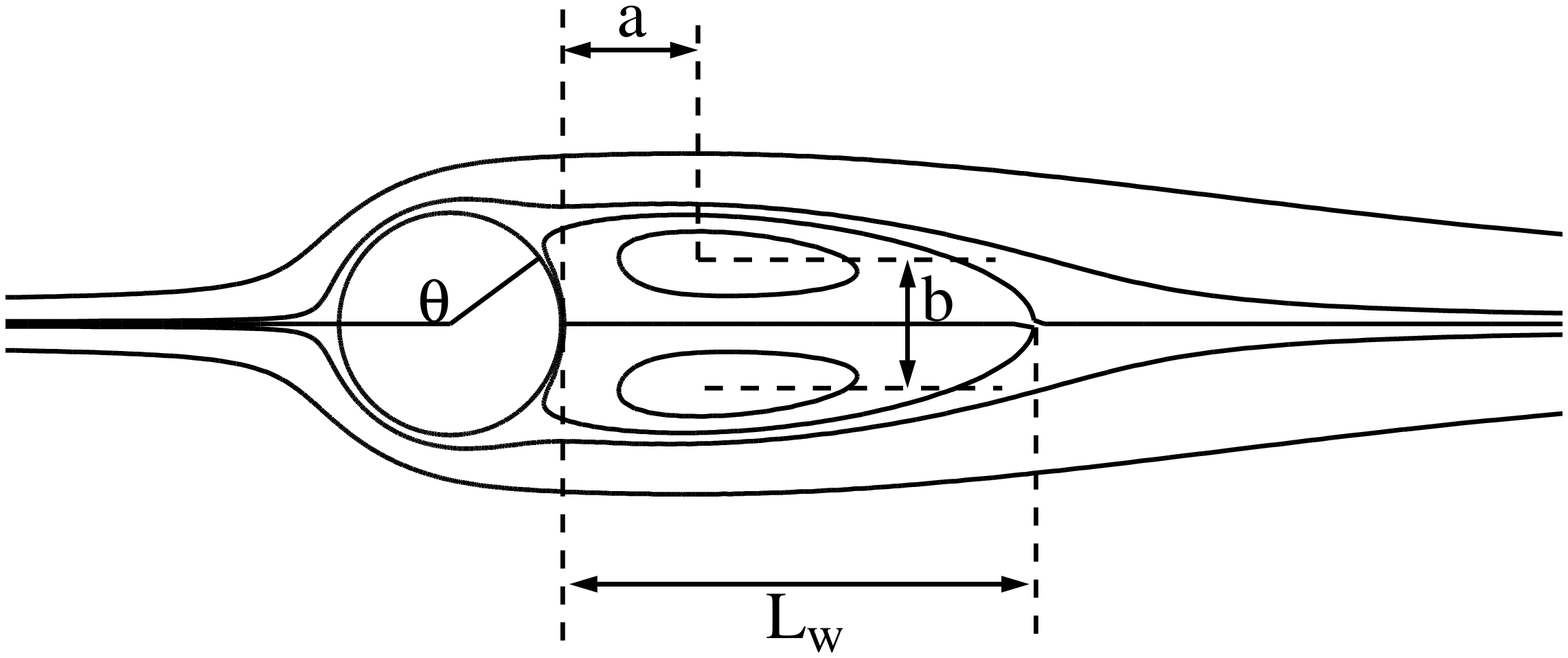}
\end{center}
\end{table}

The purpose of this study is to provide a reference solution of the cylinder flow problem at $Re=40$. The aim is not the physical investigation of this well documented flow but the achievement of high-accuracy. For that purpose, pseudo-spectral methods are used in cylindrical coordinates. In addition, to mimic satisfactory an infinite flow domain, an improved method for external boundary conditions is proposed. The solution obtained is made available at any location between the cylinder and computational domain boundaries thanks to a spectral interpolation. The full access to the resulting database is given through a short user-friendly script file that runs using the software GNU Octave or Matlab, so that the solution can be known at any specific location with a preserved accuracy. This facility is found to allow the validation of any numerical code with a possible convergence study, especially if the reference solution is used as boundary conditions. 

To illustrate the potential benefit offered by this solution, two different numerical strategies are assessed with a specific analysis of their numerical errors. These two strategies are based on the same numerical schemes of standard accuracy. The first approach corresponds to the body-fitted strategy for which a cylindrical mesh is used. The second approach is based on an Immersed Boundary Method (IBM, see Mittal \& Iaccarino \cite{mittal&iaccarino05} for an overview) based on a Cartesian mesh. In fact, the improvement of IBM was the initial motivation of this work. This type of body treatment is mainly empirical due to the difficulty of a rigorous numerical analysis of the solution convergence regarding the forcing and its numerical implementation. In this context, the availability of a reference solution, physically realistic and challenging, could be highly beneficial for further development in this area. Here, the standard direct forcing method is assessed by considering two different levels of accuracy for the near-body treatment. Because both the forcing and the numerical schemes used are standard, the resulting errors exhibited in this paper can be used as reference to conclude about the relative accuracy of any other combinations of IBM and numerical code.

The organisation of the paper is as follows. In section \ref{num}, the numerical methods used to compute the reference solution are presented with some details about the sensitivity of results. An estimation of the accuracy of the reference solution is provided in section \ref{estrefsol}. Sections \ref{macyl} and \ref{macart} are dedicated to code validation based on the reference solution for a body-fitted and an IBM approach respectively. Finally, the contribution of this work is summarized in section \ref{conclusion} where some recommendations can be found. To make very easy the access to the reference solution for the user interested in, a brief explanation of the use of the  Gnu Octave/Matlab file script is given in Appendix A.

\section{Numerical methods and reference solution\label{num}}

%% Numerical method:
The incompressible Navier-Stokes equations are solved with a pseudo-spectral code in cylindrical coordinates. In the azimuthal direction ($\theta$) the solution is expanded in Fourier series with $N_\theta$ modes. To avoid the aliasing effect the 3/2 Orzag's rules is applied. The Chebyshev collocation method is used in radial direction ($r$) with $N_r$ nodes. The Gauss-Lobatto-Chebyshev grid $\xi$ is expanded from the interval $[-1,~ 1]$ to the physical domain $[1/2,~ r_\infty]$ through a simple mapping : $r = 1/2 + r_\infty(1-\xi)/2$. A more complex mapping, with a refined mesh only near the cylinder, has been attempted but the results are poorly satisfactory because of the deterioration of the condition number of differentiation matrix. The time marching is second order accurate, implicit in time for the viscous term, with a Chorin-Temam \cite{chorin68,temam69} prediction projection scheme for the divergence free condition. The pressure is approximated by polynomials of two units lower-order than for the velocity, in order to compute a pressure unpolluted by spurious modes \cite{Botella} with only one collocation grid. An alternative method would be coupling between pressure and velocity through traction boundary conditions \cite{pouxetal11}.

%% The boundary conditions issue
The main difficulty encountered in obtaining accurate numerical solution is the satisfactory treatment of boundary conditions at large distance from the cylinder. Although a physically non confined cylinder is viewed, the discretized equations require finite boundaries in the far field. The lateral extensions of domain boundaries are responsible for the ubiquitous blockage effects, characterized by the ratio between the lateral boundary position and the diameter of the cylinder. On the other hand the outflow boundary conditions, downstream the cylinder, are applied in order to allow phenomena generated in the domain of interest to pass through the synthetic boundary without undergoing significant distortion and without influencing the interior solution.

The accuracy depends on the three following parameters: the resolution (number of grid points), the extension of the computational domain, and the boundary condition treatment.

No slip boundary condition is imposed on the cylinder. In the far field, the selection of proper boundary condition is a critical issue for flow simulation in unbounded domain. For example Pradeep and Hussain \cite{Pradeep} have shown that an artificial zero circulation constraint forces an unphysical instability of an isolated vortex. For body wake flows, the numerical domain must extend to a large distance downstream of the body. Boundary conditions like periodicity, convective, Dirichlet or Neumann are often used but not fully satisfying. In conjunction the exterior boundaries are generally placed very far from the cylinder surface to avoid unphysical influence from the boundary conditions.

In the present paper we extend a recent procedure proposed by Hasan \textit{et al.} \cite{Hasan} to extrapolate velocities at the outflow boundary. The radial variation of the velocity field at large distances from the rigid body is assumed to be proportional to $1/r^2$, which can be inferred from mass conservation and vorticity considerations. In fact the proposed boundary condition is not completely physically consistent because it implicitly assumes a irrotational outflow condition. For rotational flow, this algebraic behaviour is not satisfied, for example the far field centreline velocity is proportional to $1/\sqrt{r}$. The outer boundary could be divided into a wake region and a region outside the wake but the domains reconnection induces a strong gradient resulting in a loss of the spectral accuracy. 

The present method involves the triple decomposition $\textbf{u} = \textbf{u}_\infty + \textbf{u}_1 + \textbf{u}_2$ where $\textbf{u}_\infty$ is the uniform streamwise velocity and the two other terms are the corrections to take into account the asymptotic wake flow and the mass momentum respectively. The first correction ($\textbf{u}_1$) is given by the Schlichting self-similar far field solution \cite{Schlichting} for a plane wake flow in the streamwise ($\mathbf{e_x}$) and vertical ($\mathbf{e_y}$) directions with
$$\textbf{u}_1\cdot\mathbf{e_x} = 1-\frac{C_D\sqrt{Re}}{4\sqrt{\pi}}\frac{e^{ -Re~r (1-\cos\theta)/2}}{\sqrt{r}} $$
$$\textbf{u}_1\cdot\mathbf{e_y} =  -\frac{\sin\theta}{2}\frac{C_D\sqrt{Re}}{4\sqrt{\pi}}\frac{e^{ -Re~r (1-\cos\theta)/2}}{\sqrt{r}} $$
\noindent where the solution has been expressed in radial and azimuthal coordinates with the hypothesis $\theta\ll 1$. Nonetheless, this solution does not satisfy the mass conservation along the boundary $r=r_\infty$. Then we apply the second correction ($\textbf{u}_2$), proposed by Hasan \textit{et al.}, which is based on the expected radial variation of velocity field at large distance from the rigid body. That asymptotic behaviour, for the radial $v_2$ and azimuthal $w_2$ velocity components, can be inferred from the mass and the circulation conservation:
$$\int_0^{2\pi} v ~ rd\theta =0 \Rightarrow \int_0^{2\pi} v_2 ~ rd\theta = -\int_0^{2\pi}v_\infty+v_1 ~ rd\theta =cste$$
$$\int_0^{2\pi} w ~ rd\theta =0 \Rightarrow \int_0^{2\pi} w_2 ~ rd\theta = -\int_0^{2\pi}w_\infty+w_1 ~ rd\theta =0$$

Hence it can be inferred that the velocity components $v_2$ and $w_2$ behave as $1/r$ and $1/r^2$ respectively, for large radius $r$.
It is then possible to extrapolate the values on the external boundary from the last point in the interior domain, namely $r_{N_r-1}$:
$$v_2 = \frac{r_{N_r-1}}{r_\infty} ~  (v-v_\infty-v_1 )_{r=r_{N_r-1}},$$
$$w_2 = \left(\frac{r_{N_r-1}}{r_\infty}\right)^2~ (w-w_\infty-w_1)_{r=r_{N_r-1}}.$$

%% Results
As previously stated, the accuracy depends on the three following parameters: the resolution (number of grid points), the extension of the computational domain, and the boundary condition treatment. The different methods are compared according to two characteristics: the sensitivity of some quantitative data with respect to the spatial discretization and the downstream evolution of the centreline velocity, which must be monotonic.

Two grids have been used, a coarse grid with $N_r\times N_\theta =100\times 512$ points and a finer grid with $N_r\times N_\theta=200\times 1024$ points. In all simulations, the divergence of the velocity field is below $10^{-14}$. The infinite norm of the velocity difference between two time-steps is below $10^{-12}$ after convergence toward the steady state. The expected flow being symmetric along the streamwise-direction, the radial velocity must be real while the azimuthal component must be purely imaginary. The numerical solution satisfies these conditions with an error below $10^{-15}$.

\begin{figure}%[ht]
\begin{center}
\includegraphics[width=8cm]{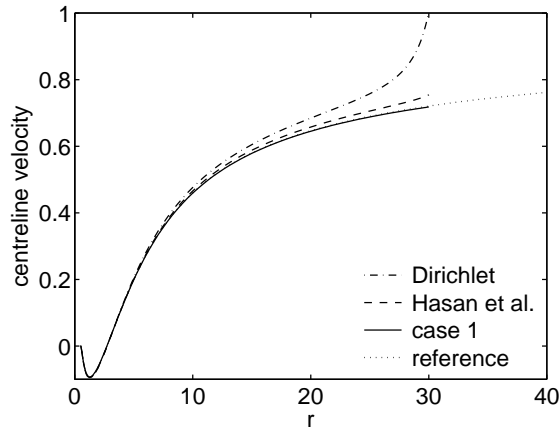}
\caption{\label{Vcl} Comparisons of the centreline velocities, simulated on the coarse grid, for different outflow boundary conditions. The dash-dotted line represents the case with the inviscid solution imposed as Dirichlet boundary condition. The dashed line represents results obtained with the extrapolated velocity as proposed by Hasan \textit{et al.}. The present results are plotted with continuous line while the reference case (\textit{i.e.} fine grid with $r_\infty=40$) is plotted in dotted line.}
\end{center}\end{figure}

As stated before the asymptotic increase of the centreline velocity is sensitive to the method used. Figure \ref{Vcl} compares the present method with Dirichlet boundary conditions and with the method introduced by Hasan \textit{et al.} \cite{Hasan}. The last two cases show a spurious inflectional point, induced by a spurious acceleration close to the outflow, while the present method presents the expected behavior.

As diagnostic tools the following quantities are selected: the drag coefficient, computed by applying the momentum flux conservation in the interior domain to avoid the uncertainty about the pressure on the boundaries
\begin{eqnarray}\label{formul_Cd}
C_D = 2r~\int_0^{2\pi}\Bigg\lbrace &-&v^2\cos\theta+v w\sin\theta -p\cos\theta \nonumber \\
&+&\frac{1}{Re}\left.\left[2\frac{\partial v}{\partial r}\cos\theta-\left(\frac{1}{r}\frac{\partial v}{\partial \theta} +\frac{\partial w}{\partial r}-\frac{w}{r}\right)\sin\theta \right]\right\rbrace ~d\theta\, , \nonumber\\
\end{eqnarray}
the wake length of the recirculating bubble $L_w$ measured from the rear stagnation point and the separation angle $\theta_s$ measured from the front stagnation point.
The influence of the grid refinement ($N_r,~N_\theta$) and the size of the domain ($r_\infty$) are summarized in the table \ref{tab_1}. The values for the $C_D$ are in good agreement with the result by Posdziec \& R. Grundmann \cite{Posdziec} $C_D=1.4942$, obtained with a computational domain up to $4000$ of cylinder diameters. 
\begin{table}[ht!]
 \caption{\label{tab_1} Drag coefficient $C_D$, nondimensional wake length $L_w/D$ and separation angle $\theta_s$ at $Re=40$, 
for various grid resolutions and domain sizes $r_\infty$. The reference case retained corresponds to the fine grid with $r_\infty=40$.}
\begin{center}
\begin{tabular}{c|ccc|ccc}
 \hline
			&     \multicolumn{3}{c|}{coarse grid}                  & \multicolumn{3}{c}{fine grid} \\
			&     \multicolumn{3}{c|}{$N_r=100$, $N_\theta=512$}    & \multicolumn{3}{c}{$N_r=200$, $N_\theta=1024$} \\
 \hline
 $r_\infty$ &  30  & 40 & 50 & 30 & 40 & 50  \\
      $C_D$ & 1.4903 & 1.5153 & 1.5024 & 1.4906 & 1.4931 & 1.4943  \\
    $L_w/D$ & 2.2340 & 2.2628 & 2.2507 & 2.2346 & 2.2360 & 2.2369   \\
 $\theta_s$ & 126.4117 & 126.2627 & 126.3315 & 126.4059 & 126.3945 & 126.3888   \\
 \hline
\end{tabular}
\end{center}
\end{table}

\begin{figure}%[ht]
\begin{center}
\includegraphics[width=0.49\textwidth]{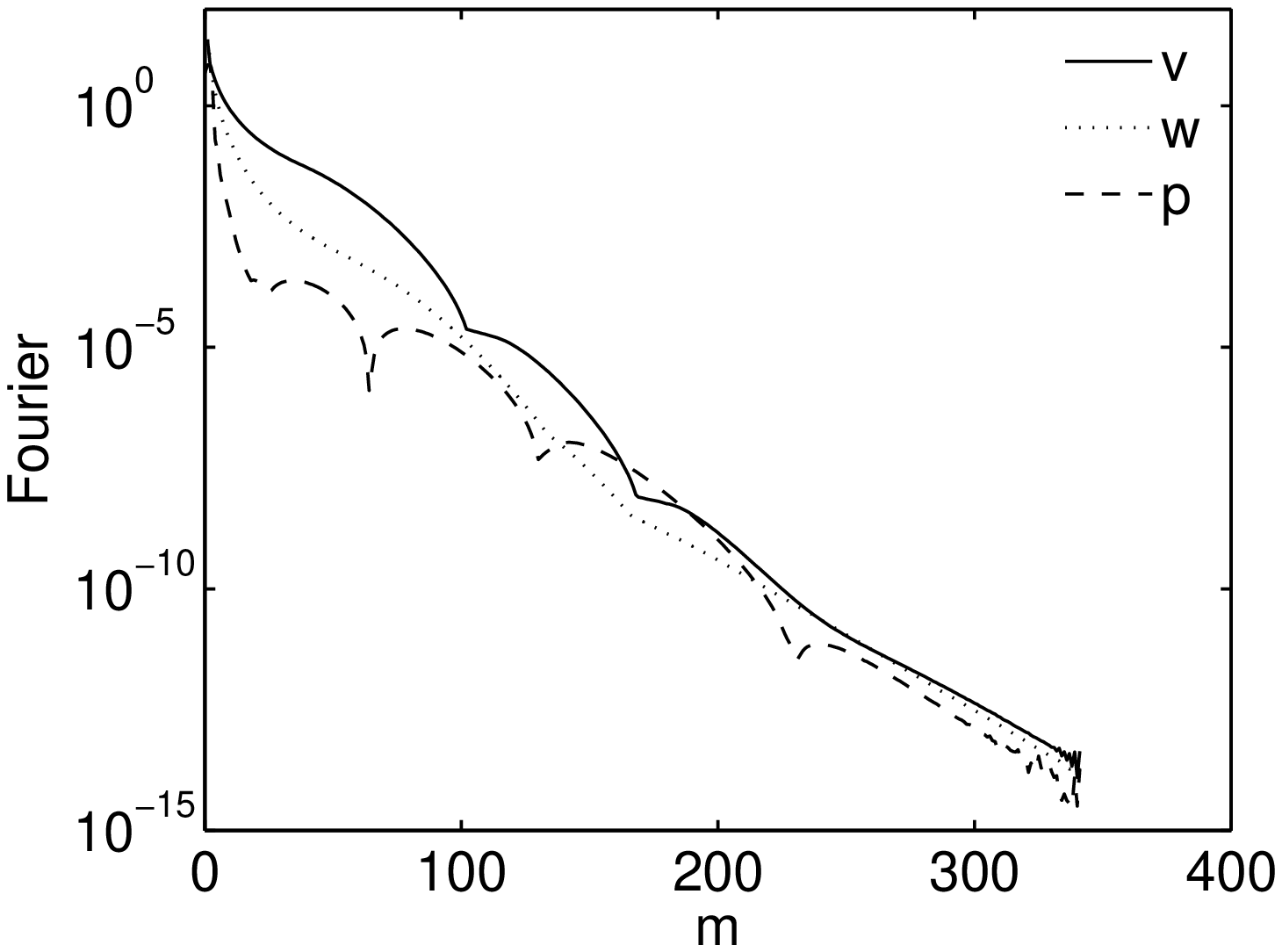}
\includegraphics[width=0.49\textwidth]{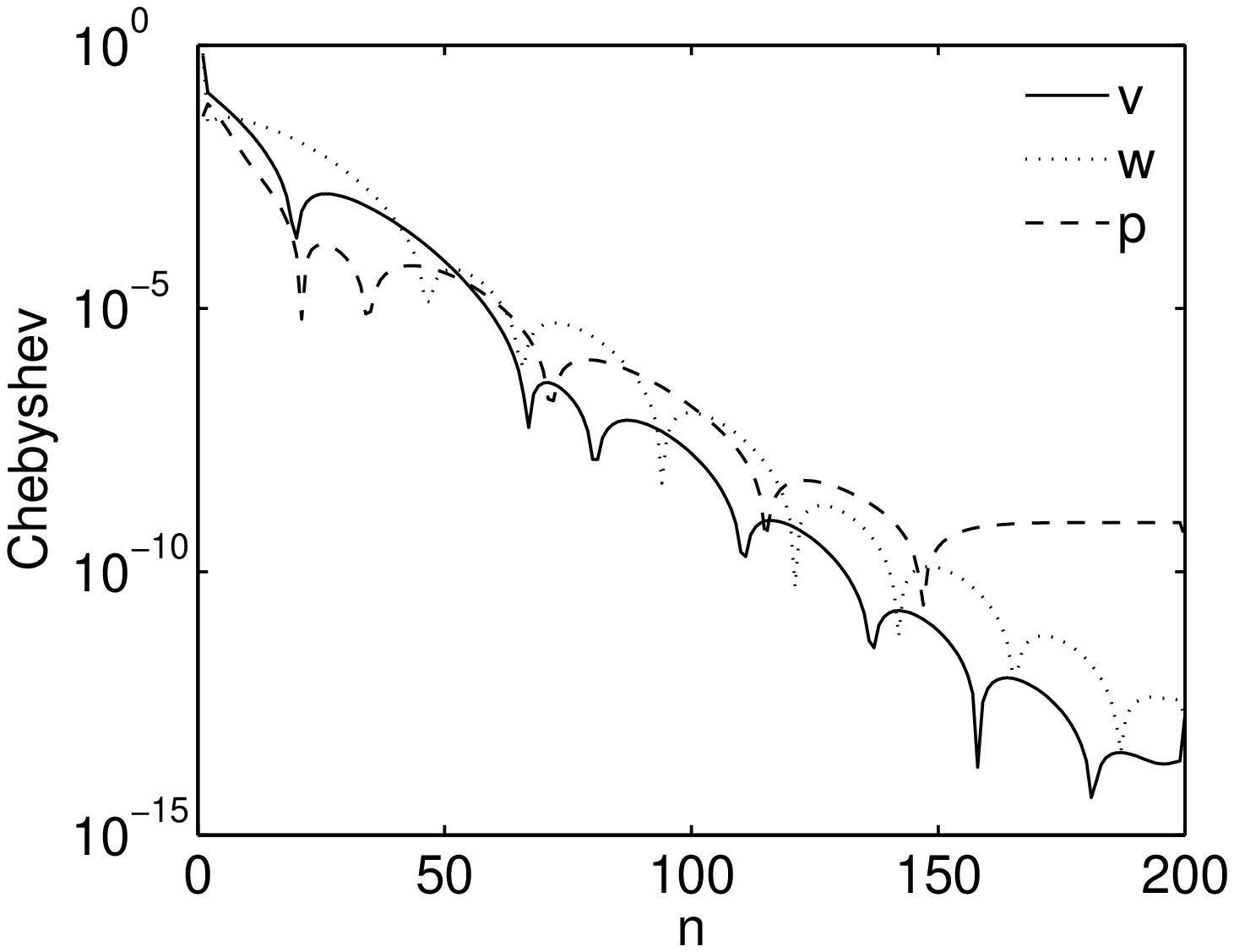}
\caption{\label{spectra} Fourier (left) and Chebyshev (right) spectra, for the radial ($v$), the azimutal ($w$) velocity components and for the pressure ($p$).}
\end{center}
\end{figure}

For the reference case, the numerical parameters retained are: $N_r=200$, $N_\theta=1024$ and $r_\infty=40$. The physical parameters obtained for this reference solution are reported in table \ref{tab_2}. The numerical accuracy is checked with the Fourier and Chebyshev spectra displayed in figure \ref{spectra}. 

The trail of the spectrum is close to the machine precision, thus the solution can be considered as numerically converged. It is worth noting that the numerical convergence interpreted through this decrease of the spectral coefficients does not necessarily correspond to the convergence toward the exact solution of the problem. The residual mismatch is due to the outer boundary conditions that cannot mimic perfectly an infinite domain.

In fact, the cylinder wake problem presents two difficulties: the discretization of the spatial differential operators and the position of the far-field boundary condition. Thus, the definition of convergence is not unique and the meshing convergence has to be distinguished from the asymptotic boundary condition effects. For extrapolated boundary conditions, both are related, since the outflow boundary conditions depends on the drag coefficient, depending itself on the spatial resolution. Even with fixed domain length, this coupling cannot ensure the convergence for the integral quantities. This phenomenon can be observed in table \ref{tab_1} where the reported values are still sensitive to the spatial resolution and to the domain size. Despite the improvement offered by the present set of outer boundary conditions, a spectral convergence of this data would need the use of an extremely extended computational domain. This condition would require a huge computational effort while being penalizing for the final convenience of the reference solution. However, for validation purpose, by using the reference solution as Dirichlet boundary condition, the two convergence properties (mesh and domain size) become uncoupled so that convergence can be expected, with the mesh refinement, for all quantities, local and global. This point will be shown in the next section.

\section{Estimation of the reference solution accuracy\label{estrefsol}}

The goal of this section is to estimate the accuracy of the reference solution when used as a diagnostic tool to check the numerical convergence. For that purpose, the same spectral code is used but with a smaller domain $r_\infty=10$. The boundary condition is spectrally interpolated from the reference solution (see Appendix A). Then, as discussed in the previous section, the convergence is not affected by the approximation of the outer boundary conditions. The grid refinement is based on the number of radial points ($N_r$). The azimuthal grid is adapted using the empirical rule $N_\theta=5 N_r$, providing an homogeneous error in both spatial directions.

Various classical norms have been used ($L_1, L_2$ and $L_\infty$). Since all of these norms lead to the same conclusion, the following analysis is restricted to the norm defined as
\begin{equation}\label{form_err}
q_\mathrm{err}= \sqrt{\frac{1}{S} \int_S (q_\mathrm{cal}-q_\mathrm{ref})^2~ dS}
\end{equation}
\noindent where $S$ is the surface of the computational domain. $q_\mathrm{cal}$ and $q_\mathrm{ref}$ stand for the calculated and reference solution respectively where $q$ can be the velocity norm $\sqrt{v^2+w^2}$ or the pressure $p$. This definition quantifies only the error on the velocity modulus, which is slightly different from the norm of the error that includes the contribution of the angular error. 

\begin{figure}
\begin{center}
\includegraphics[width=8cm]{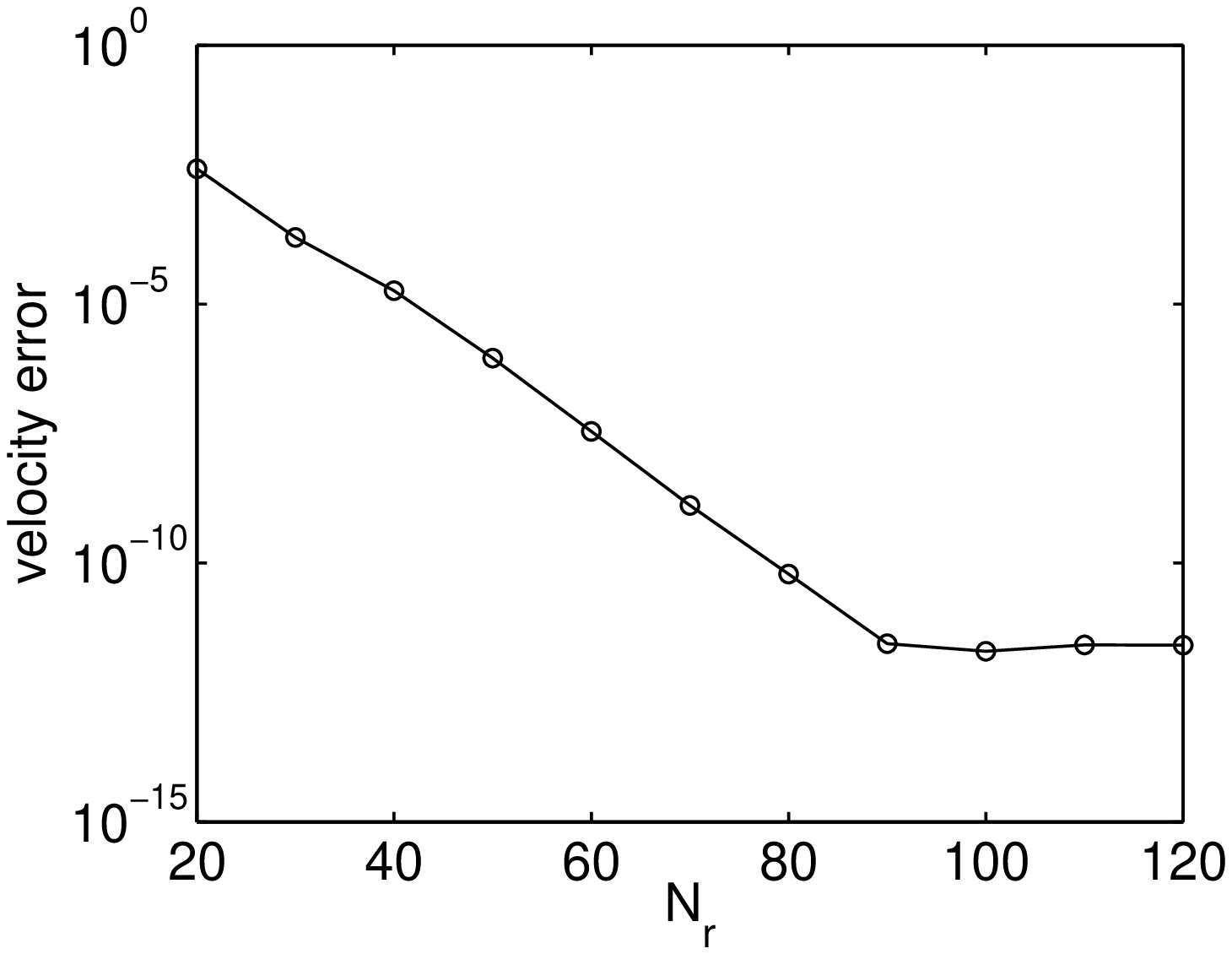} \\
\includegraphics[width=8cm]{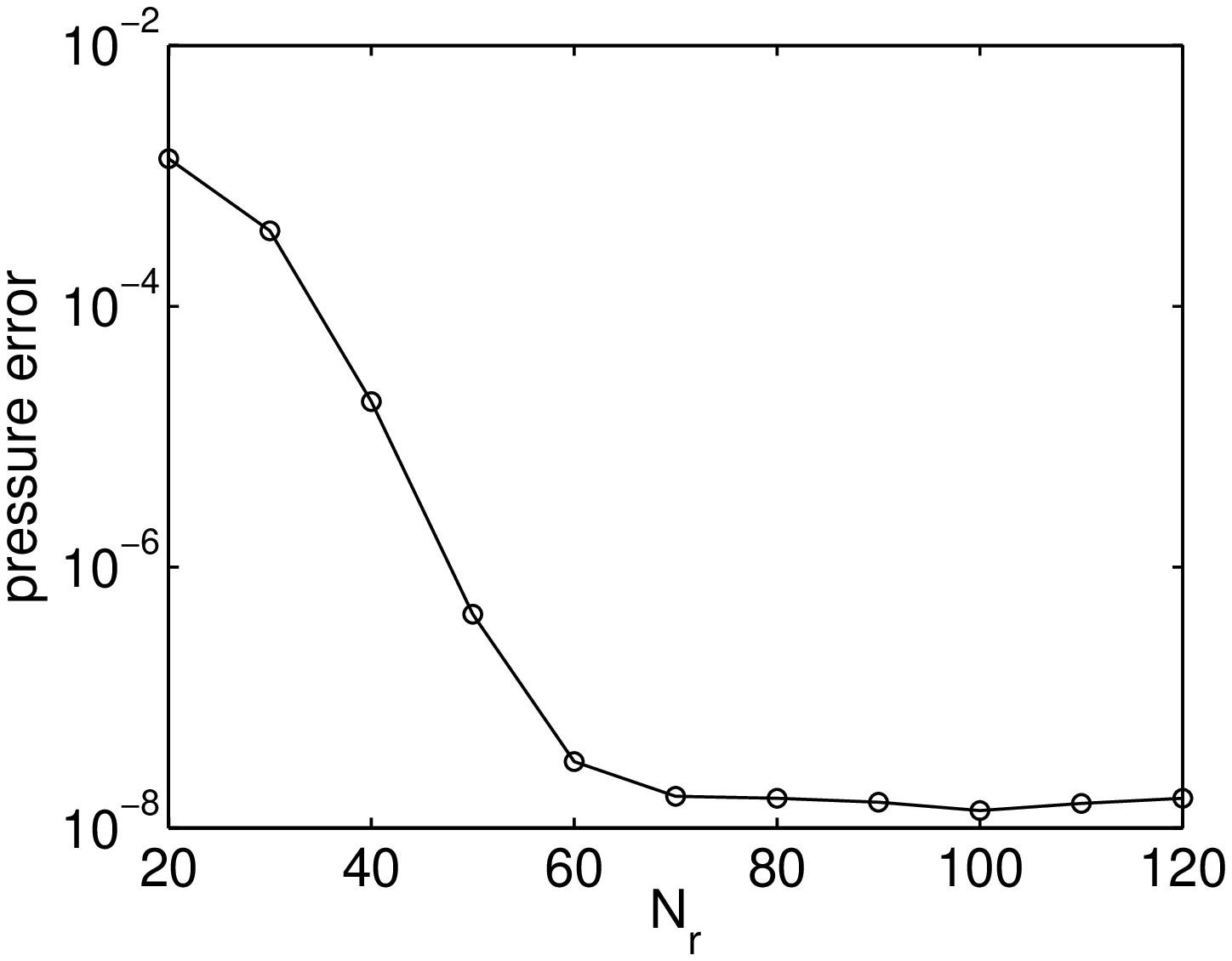} \\
\includegraphics[width=8cm]{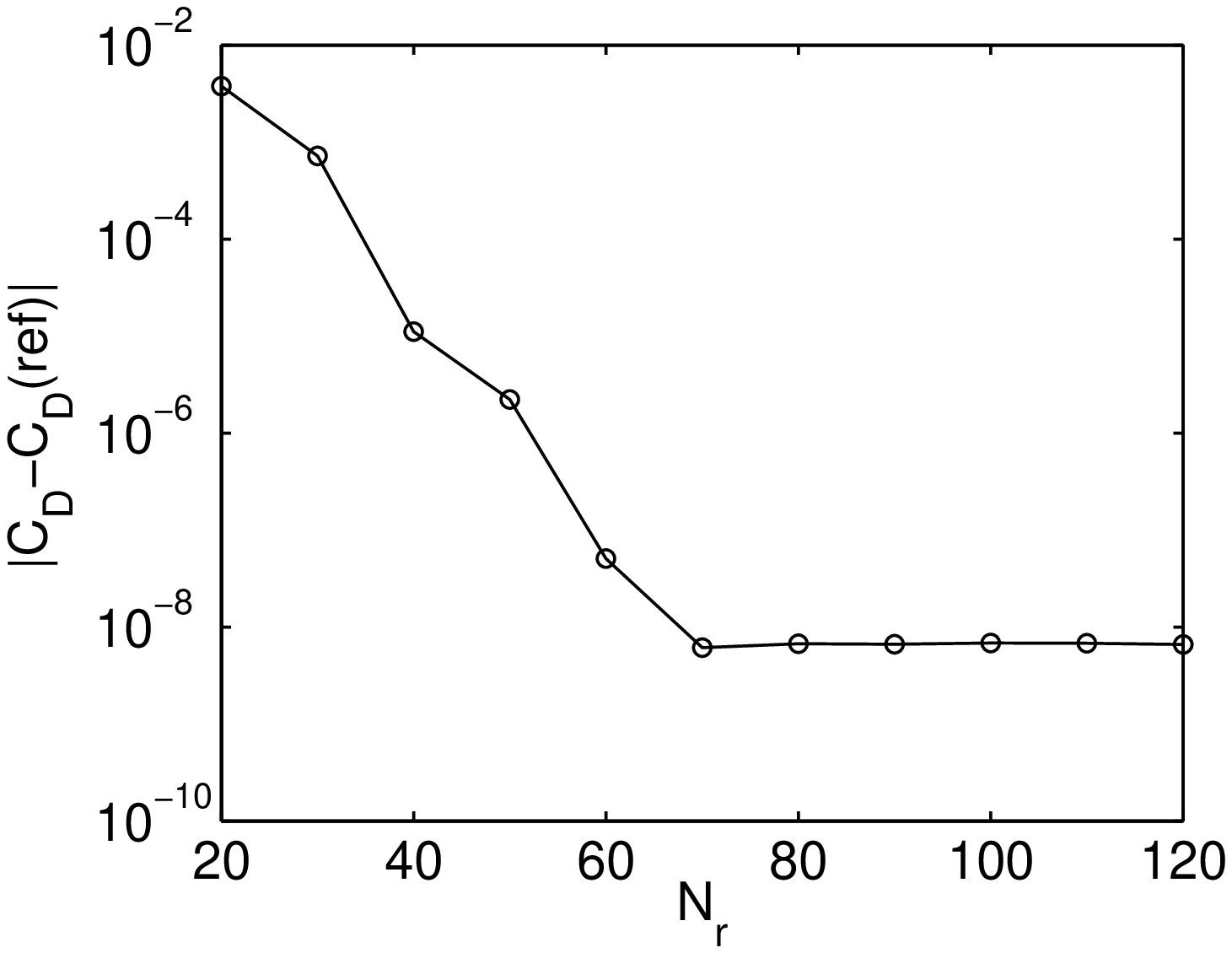}
\caption{\label{converspect}Numerical convergence for the velocity, pressure and drag coefficients (the pressure constant value is taken at $r=0.5^+$ and $\theta=0$).}
\end{center}
\end{figure}

The convergence plots are shown in figure \ref{converspect}. 

For the velocity field, a spectral (exponential) convergence is observed until the value $2.75\, 10^{-12}$, reached with $N_r=90$. For the pressure the evolution is not exponential and saturates at the value $2.47 \, 10^{-8}$ for $N_r=70$. This behaviour can be explained by the absence of boundary conditions for the pressure. This lack creates fundamental difficulties for the projection scheme during simulations, also for the spectral interpolation in the radial direction for the reference solution. This error is passed to the drag coefficient, computed in the interior domain by the formula (\ref{formul_Cd}).

The saturation of the convergence provides an estimation of the accuracy of the reference solution up to 11 digits for the velocity and 7 digits for the pressure. This accuracy is by far enough for the validation of a numerical code using typical resolutions. However, for an asymptotic convergence at very high resolution, the threesholds of $10^{-11}$ and $10^{-7}$ as to be considered as the limits of the reference solution.

\section{Test 1: Validation of a MAC cylindrical code\label{macyl}}

As a first step, the convergence obtained on a cylindrical grid using a finite-difference code of second order accuracy is considered. The mesh is fully staggered as initially proposed by Harlow \& Welch \cite{harlow&welch65} in their Marker and Cell (MAC) method. The equations actually solved are a specific version of the incompressible Navier-Stokes equations where the unknown variables are $(q_r,q_\theta)=(r u_r,r u_\theta)$ with $(r,\theta)$ and $(u_r,u_\theta)$ the cylindrical coordinates and velocity components respectively. The use of these variables overcomes the singularity problem at $r=0$ in the governing equations as discussed by Orlandi \cite{orlandi00}. For the present flow configuration, this utility has not to be used. The incompressibility condition is ensured up to the machine accuracy using the direct Poisson solver FishPack\footnote{http://www2.cisl.ucar.edu/resources/legacy/fishpack}. Because the grid is cylindrical, no-slip boundary conditions can be straightforwardly applied on the cylinder surface so that the present test is an example of body-fitted approach.

Four levels of spatial resolution are considered using the mesh node numbers $n_r\times n_\theta=91\times 360$, $181 \times 720$, $361 \times 1440$ and $721 \times 2880$. The computation domain is given by $0.5D\leq r\leq 10.5D$ so that the mesh size is given by $\Delta r = 10D/(n_r-1)$ and $\Delta \theta = 2\pi/n_\theta$. The resolution in the azimuthal direction is chosen to match the number of degrees of freedom using a Cartesian grid with an IBM (see next section). 

The boundary condition at $r=10.5D$ is imposed as a Dirichlet condition on the velocity using the reference solution. As discussed in the previous section, this treatment avoids the introduction of additional numerical errors associated with the outer boundary conditions, allowing us to focus on those associated with the discretization.

The flow over a cylinder at $Re=40$ is computed until a steady state is reached. It has been checked that initial conditions only modify the transient stage toward the steady flow but not the computed final state itself. 

Figure \ref{convmacyl} presents the convergence of the solution for the velocity norm $|u|=\sqrt{u_r^2+u_\theta^2}$ and the pressure $p$. 

\begin{figure}
\begin{center}
\includegraphics[bb=50 80 554 760,height=0.49\textwidth,angle=-90,clip=]{conv_U_MAC_cylindrique.ps}
\includegraphics[bb=50 80 554 760,height=0.49\textwidth,angle=-90,clip=]{conv_P_MAC_cylindrique.ps}
\end{center}
\unitlength\columnwidth
\vspace{-0.5\baselineskip}
\begin{picture}(1,0)
\put(0.2,0){$\sqrt{n_r\times n_\theta}$}
\put(0.69,0){$\sqrt{n_r\times n_\theta}$}
\end{picture}
\caption{Error decrease with the spatial resolution. Left: velocity error $|u|_\mathrm{err}$. Right: pressure error $p_\mathrm{err}$  (the pressure constant value is taken as averaged value on the computational domain).}
\label{convmacyl}
\end{figure}

Here, the convergence is examined through the error expressed as in (\ref{form_err}) with
\begin{equation}
q_\mathrm{err} = \sqrt{
\frac{\Delta r \Delta \theta}{110\pi D^2}
\sum_{i=1}^{n_r-1} \sum_{j=1}^{n_\theta} 
\left[q_\mathrm{cal}(r_i,\theta_j) - q_\mathrm{ref}(r_i,\theta_j)\right]^2 
r_i} 
\end{equation}
where $q_\mathrm{cal}$ and $q_\mathrm{ref}$ are respectively the calculated and reference solutions at the same location $(r_i,\theta_j)$.
Note that, for comparison purpose, the error has again been normalized with the computational domain surface. The set of locations $(r_i,\theta_j)$ depends on the spatial resolution. Due to the staggered grid, it could also depends on the variables considered among $(u_r,u_\theta,p)$. Here, for simplicity, only locations $(r_i,\theta_j)$ corresponding to pressure nodes are considered, requiring to perform a second order interpolation for the velocity components $(u_r,u_\theta)$. Thanks to spectral interpolation, the reference solution is available in any location with a similar accuracy. This property is very convenient in the present context where the computation of $q_\mathrm{err}$ can be done as easily as if the exact solution were known as reference.
 
Figure \ref{convmacyl} shows a second order convergence for the velocity and the pressure as expected regarding the numerical schemes actually used. These first results illustrate the way of using the reference solution to check a numerical code through a formal convergence test. The achievement of the expected convergence order is an additional confirmation that the present spectral solution can be used as a reference solution. Even at the highest resolution, the numerical error is by far larger than the accuracy of the reference solution reported in section \ref{estrefsol}, with a ratio of about 7 and 3 orders of magnitude for the velocity and the pressure respectively.

\begin{figure}
\begin{center}
\includegraphics[bb=110 100 450 440,width=0.44\textwidth,clip=]{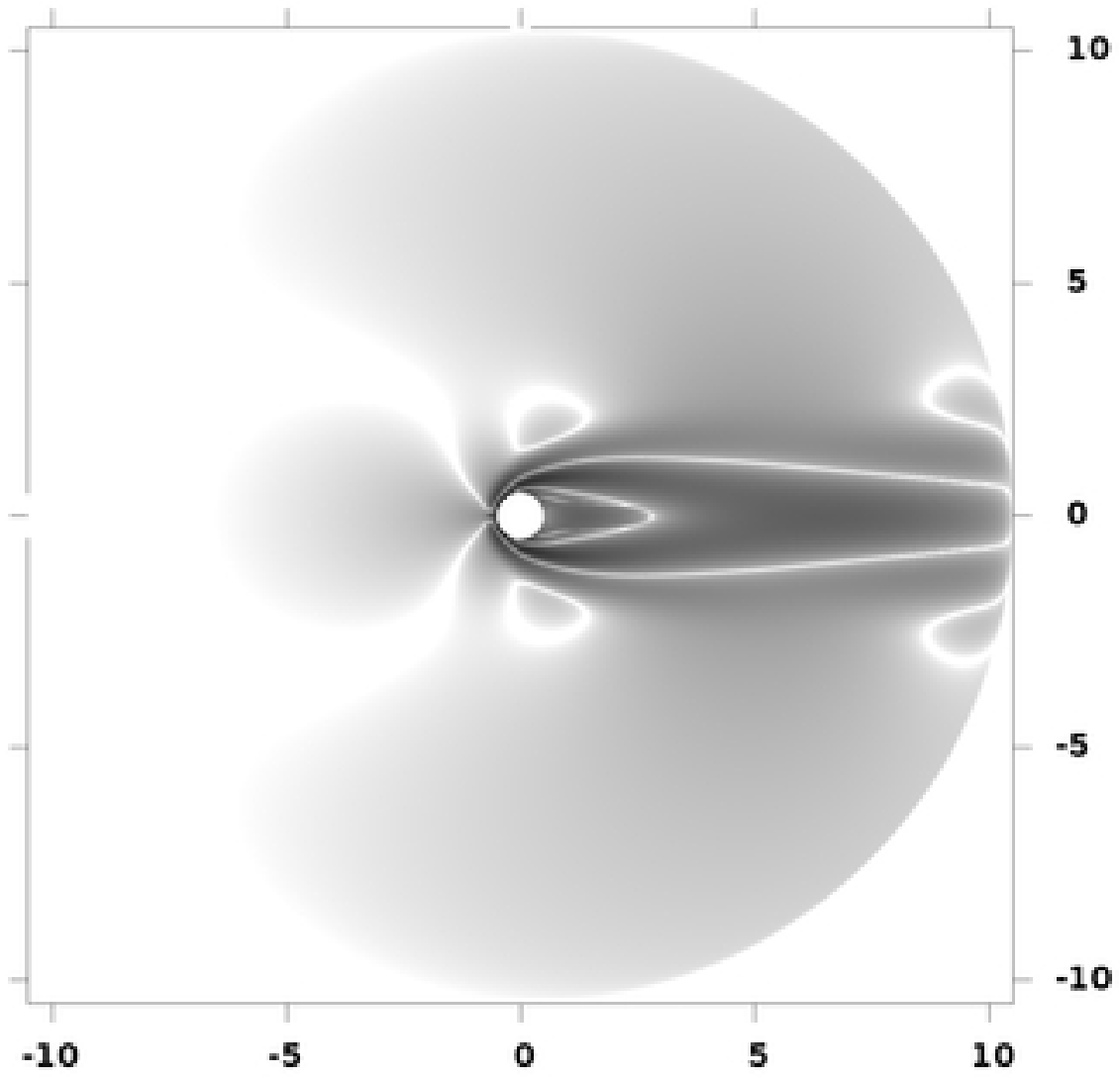}
\includegraphics[bb=110 100 450 440,width=0.44\textwidth,clip=]{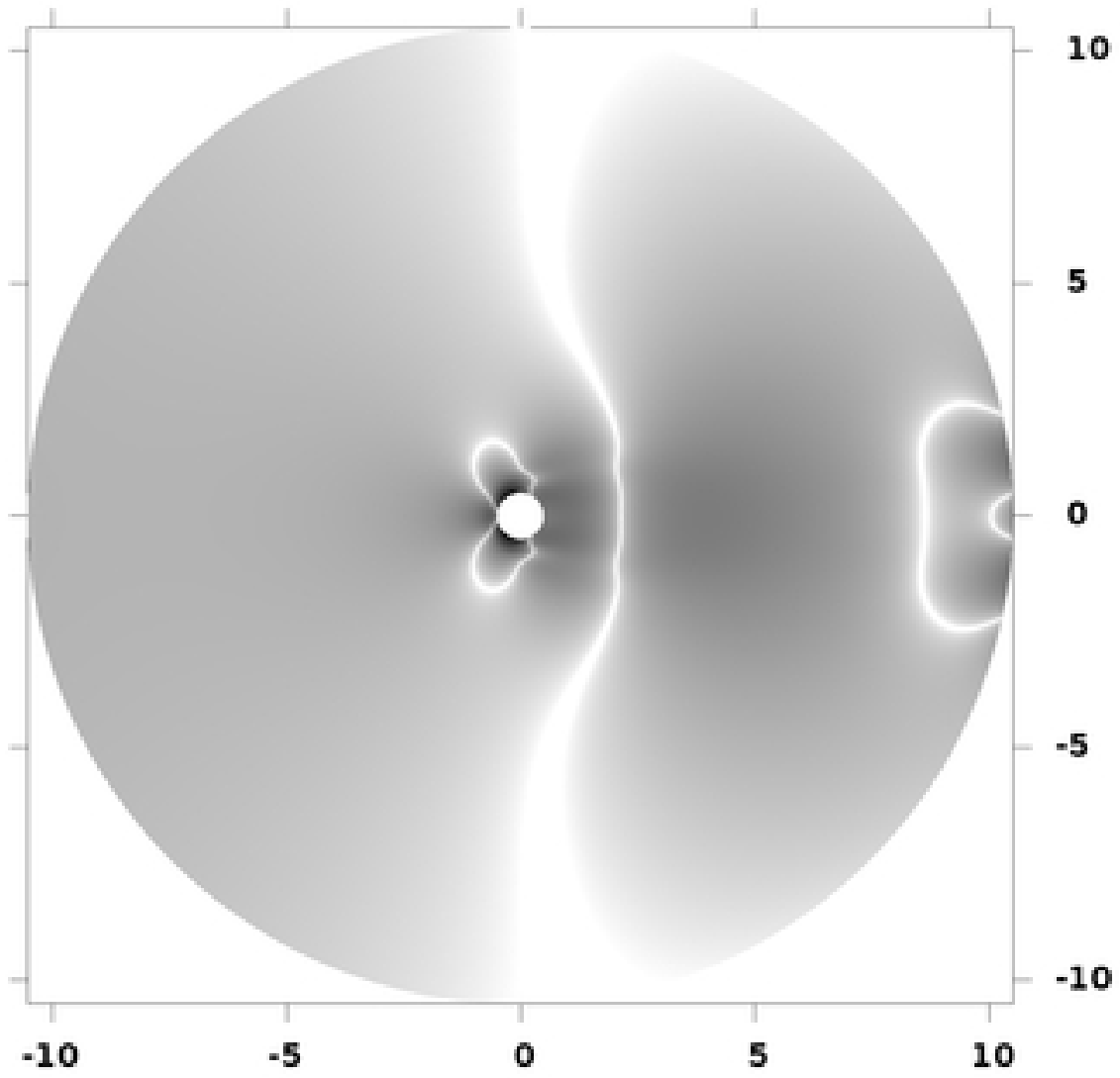}
\includegraphics[width=0.075\textwidth,clip=]{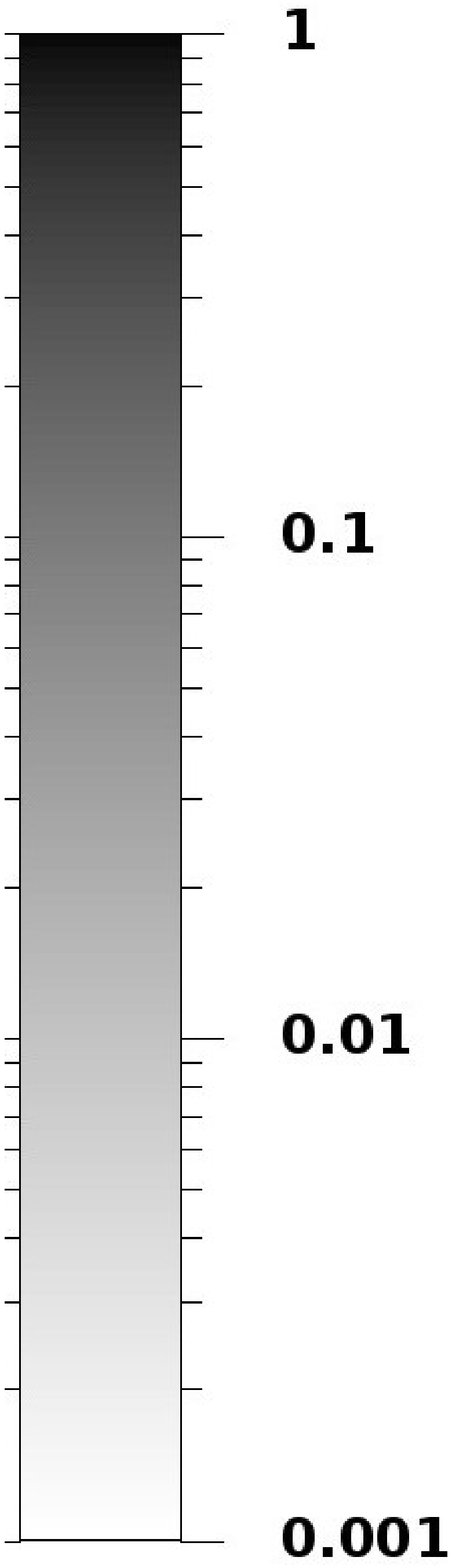}
\end{center}
\unitlength\columnwidth
\vspace{-1.5\baselineskip}
\begin{picture}(1,0)
\put(0.21,0){$x$}
\put(0.66,0){$x$}
\put(-0.005,0.23){$y$}
\end{picture}
\caption{Maps of normalized error $q_\mathrm{err}/{q_\mathrm{err}^{\max}}\in[0.001,1]$ at the highest resolution $n_r\times n_\theta=721 \times 2880$. Left: velocity error $|u|_\mathrm{err}$. Right: pressure error  $p_\mathrm{err}$.}
\label{mapmacyl}
\end{figure}

The full knowledge of the reference solution at any location also allows a direct comparison between the calculated and reference solutions by the plot of the error field $q_\mathrm{err}$. The maps of this quantity for the velocity and the pressure, shown in figure \ref{mapmacyl}, indicate the flow region where the numerical errors prevail for the highest resolution. For clarity, a normalization by the maximum value is performed and a logarithmic scale colormap is used on the range $[0.001,1]$. 

Because the mesh size increases with the distance from the cylinder for the present cylindrical mesh, significant errors are found far from the body in the wake region. However, the highest errors can be observed in the neighbourhood of the cylinder, especially slightly upstream from the two separation points where the flow is deflected by the cylinder. These near-cylinder errors are also recovered for the pressure (see figure \ref{mapmacyl}-right), with far from the body, some dominant errors located in the wake region at the boundary of the computational domain. This behaviour can be interpreted as the consequence of prescribing the velocity as a Dirichlet boundary condition that can be in conflict, for the present discretization, with the preservation of the free divergence condition up to the machine accuracy. Because the pressure considered here is the variable used to ensure the incompressibility condition (in the framework of the projection method and according to the present second-order discretization), it has to adapt to the hard velocity Dirichlet boundary condition, leading to an increase of its approximation error near the boundary of the computational domain, this approximation remaining however second order accurate as suggested by figure \ref{convmacyl}.

In terms of local error near the cylinder surface, the critical region seems to be the upstream part of the body. The resulting error pattern in the full computational domain cannot be rigorously interpreted, but it seems reasonable to consider that outside the recirculation region, the dominant convective component of the flow from upstream to downstream is susceptible to amplify the error further downstream. For the present flow, it can be conjectured that the overall error is highly sensitive to the most upstream errors. In the present case, the predominance of errors near the separations seems to be physically relevant if we consider the well-known sensitivity of the separation process, especially for smooth geometry where no-singularity can fix the flow separation. Naturally, the maps presented in figure \ref{mapmacyl} could be very different if a radial and azimuthal refinement were used to concentrate for instance mesh nodes near the separation points. In the same way, a different error pattern could be obtained using another body fitted approach based on another grid organization, structured or not. Using the present reference solutions, all these situations could be easily considered depending on the type of numerical code to validate.

\section{Test 2: Validation of a MAC Cartesian code with an IBM\label{macart}}

As a second example, the accuracy offered by an IBM is investigated in this section. A MAC discretization is again used but with a simple Cartesian mesh where the variables are the velocity components $u_x$ and $u_y$ in a computational domain $L_x\times L_y=20D\times 20D$. As in the previous section, Dirichlet boundary conditions are imposed on the velocity at $x=\pm 10D$ and $y=\pm 10D$ using the reference solution. To model the cylinder, an IBM is used through a direct forcing technique, following  Fadlun \textit{et al.} \cite{fadlunetal00}, for which body forces are used to impose $(u_x,u_y)=(0,0)$ inside the cylinder. Two different forcing are tested, a first one with no interpolation to specify the exact location of the cylinder surface and a second one where the forcing is also applied at mesh nodes just next to the immersed boundary through a second-order interpolation, see Fadlun \textit{et al.} \cite{fadlunetal00} for more details. 

Again, four different spatial resolutions are considered with $n_x\times n_y=181\times 181$, $361\times 361$, $721\times 721$ and $1441\times 1441$. These computational grids allow us to match the radial resolutions considered in the previous section ($\Delta x = \Delta y = \Delta r$) while leading to the same number of degrees of freedom. Here, the error is measured in the physical domain \textit{i.e.} outside the cylinder, as
\begin{equation}
q_\mathrm{err} = \sqrt{
\frac{\Delta x \Delta y}{(400-\pi/4)D^2}
\sum_{i=1}^{n_x-1} \sum_{j=1}^{n_y-1} 
\left[1-\varepsilon(x_i,y_j)\right]
\left[q_\mathrm{cal}(x_i,y_j) - q_\mathrm{ref}(x_i,y_j)\right]^2}
\label{l2cart}
\end{equation}
with $\varepsilon=1$ inside the cylinder and $\varepsilon=0$ in the fluid region. In addition, because the velocity norm $|u|_\mathrm{cal}$ is estimated at the pressure node location through a second-order interpolation of the velocity components $(u_x,u_y)$, the contribution of nodes involving values inside the cylinder has been cancelled (imposing $\varepsilon=1$) in equation (\ref{l2cart}). The decrease of $|u|_\mathrm{err}$ and $p_\mathrm{err}$ with the spatial resolution is presented in figure \ref{convmacart} for the two different forcing methods. 

\begin{figure}
\begin{center}
\includegraphics[bb=50 80 554 760,height=0.49\textwidth,angle=-90,clip=]{conv_U_MAC_cartesien_4pts.ps}
\includegraphics[bb=50 80 554 760,height=0.49\textwidth,angle=-90,clip=]{conv_P_MAC_cartesien_4pts.ps}
\end{center}
\unitlength\columnwidth
\vspace{-0.5\baselineskip}
\begin{picture}(1,0)
\put(0.2,0){$\sqrt{n_x\times n_y}$}
\put(0.69,0){$\sqrt{n_x\times n_y}$}
\end{picture}
\caption{Error decrease with the spatial resolution. Left: velocity error $|u|_\mathrm{err}$. Right: pressure error $p_\mathrm{err}$ (the pressure constant value is taken as averaged value on the computational domain).}
\label{convmacart}
\end{figure}

As expected, the case without any interpolation across the immersed boundary leads nearly to a first-order convergence for the velocity and the pressure. The clear benefit offered by the use of an interpolation can be observed in the same figure where both the velocity and pressure are better predicted. The velocity is found to approach the second order accuracy thanks to the interpolation. However, it can be observed that the pressure convergence is not as good as for the velocity, especially for the highest resolutions. In the present IBM, no specific treatment is imposed on the pressure that has to adapt to the forcing without any boundary condition prescription. This lack of control of the pressure is probably the reason of its less favourable convergence. It suggests that the pressure treatment could be the main limiting factor of the present IBM in terms of accuracy, especially for an improvement of its formal order. To our knowledge, this is the first time that these velocity and pressure convergences are shown using an IBM for a flow exhibiting separation and recirculation.

\begin{figure}
\begin{center}
\includegraphics[bb=110 100 450 440,width=0.44\textwidth,clip=]{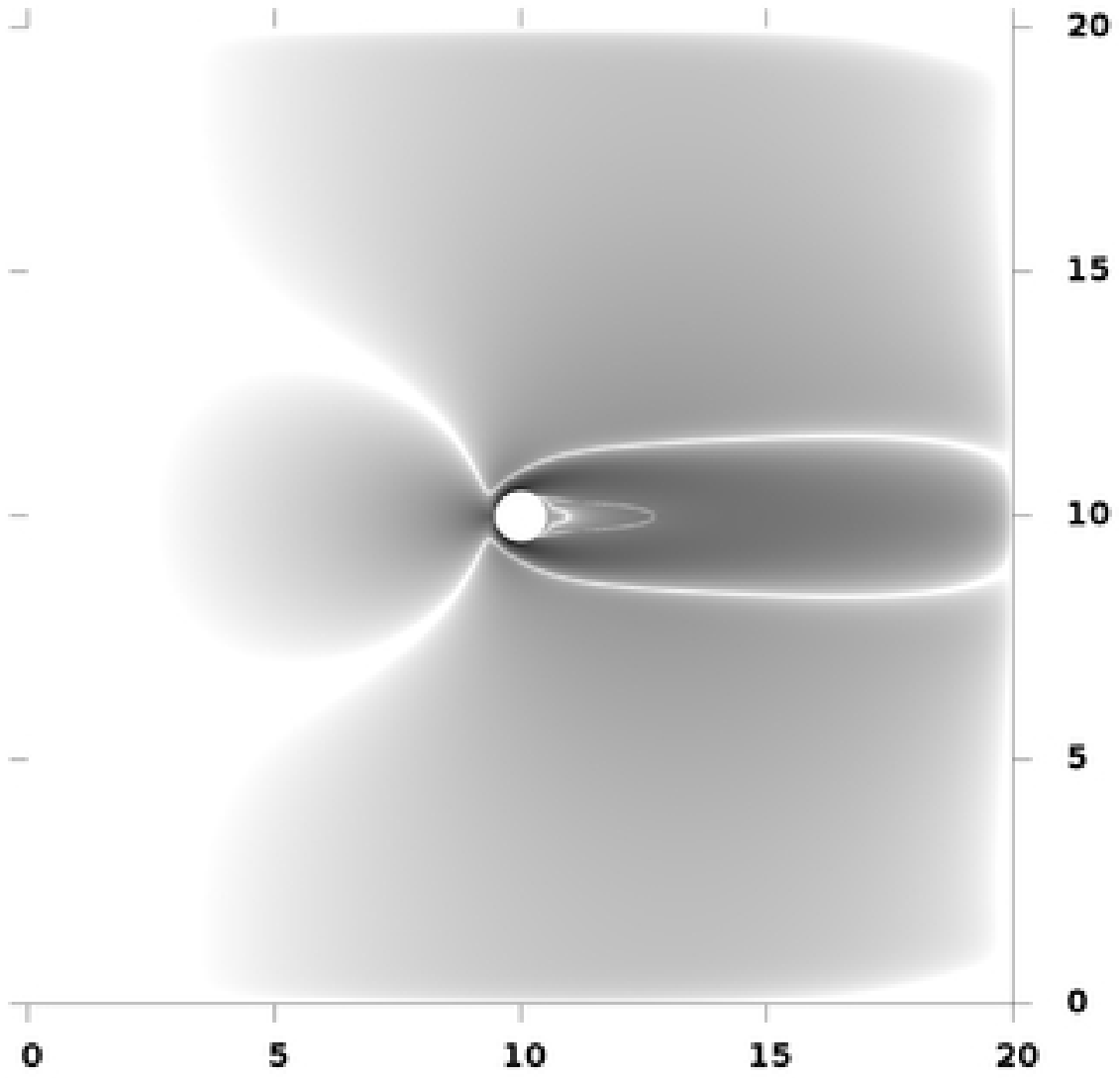}
\includegraphics[bb=110 100 450 440,width=0.44\textwidth,clip=]{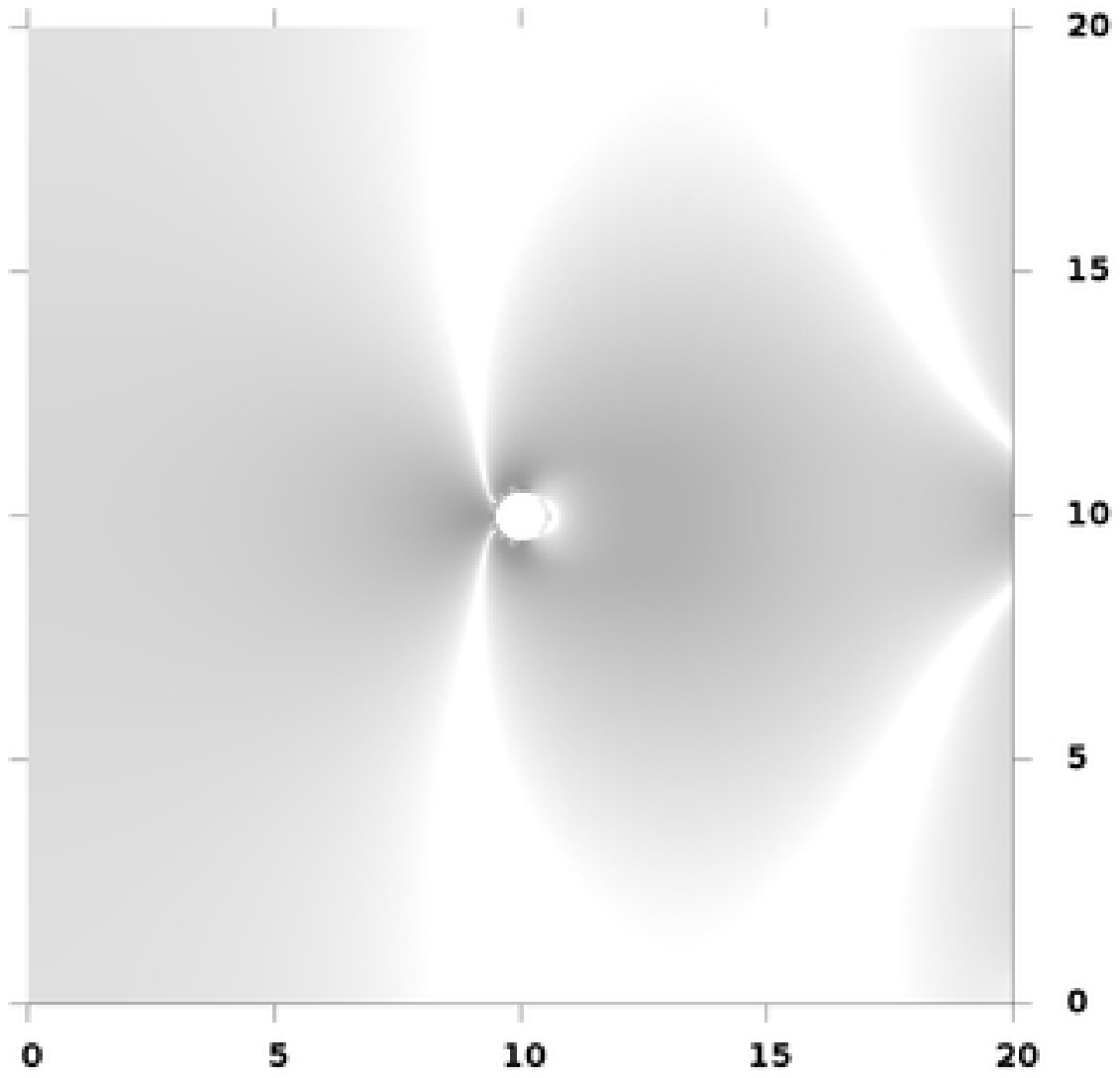}
\includegraphics[width=0.075\textwidth,clip=]{log_0001c_GRIS}
\includegraphics[bb=110 100 450 440,width=0.44\textwidth,clip=]{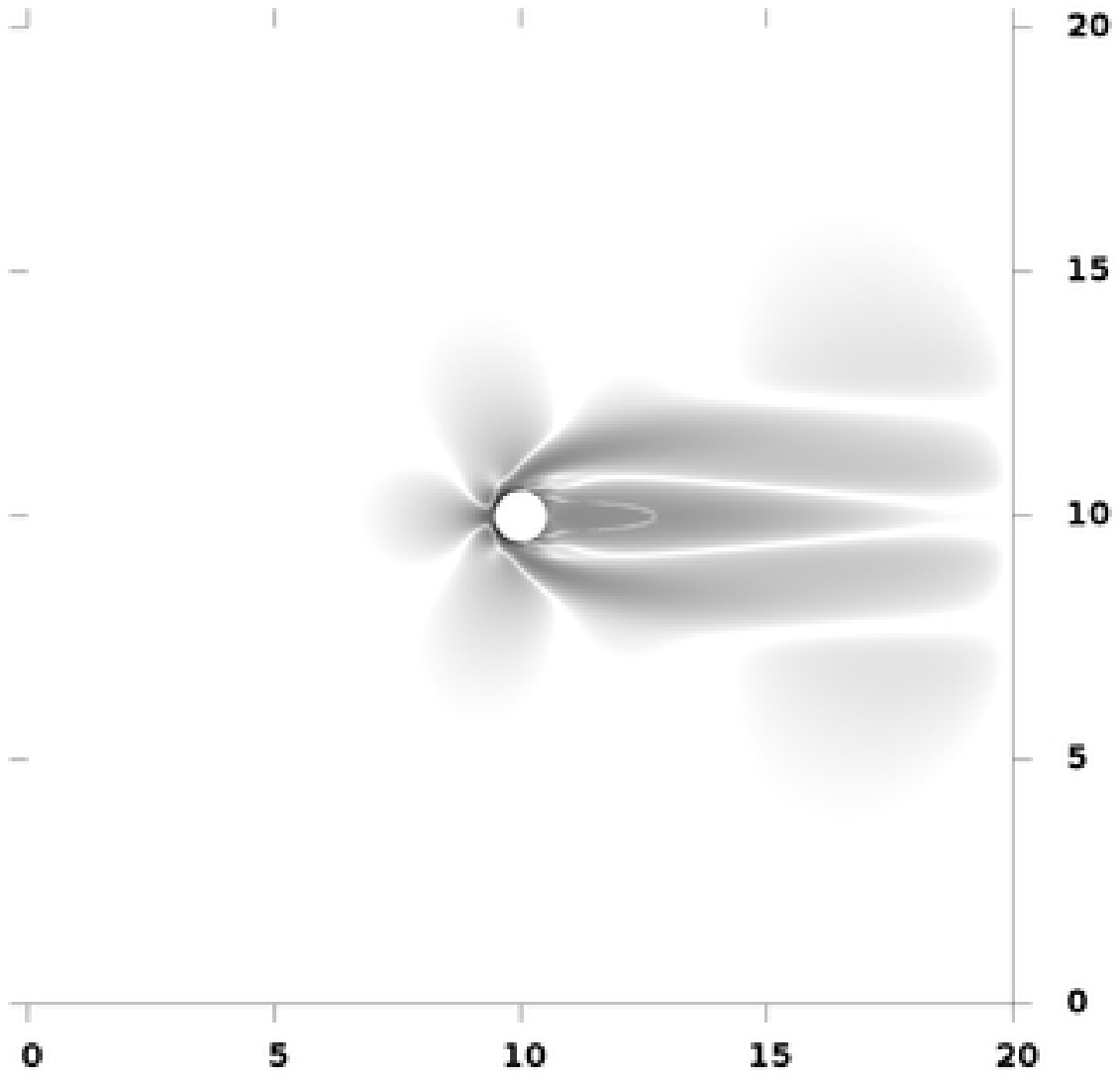}
\includegraphics[bb=110 100 450 440,width=0.44\textwidth,clip=]{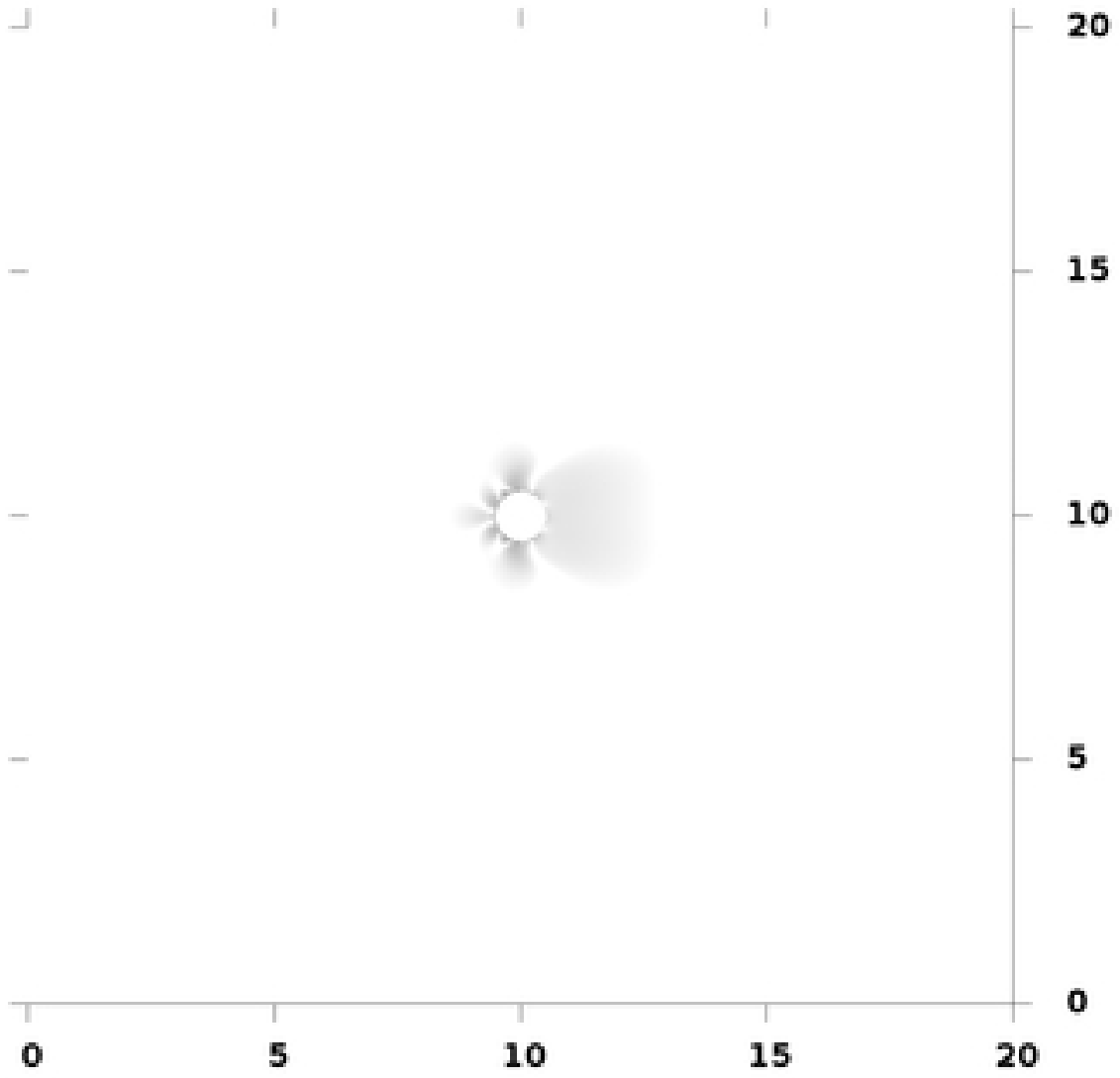}
\includegraphics[width=0.075\textwidth,clip=]{log_0001c_GRIS}
\end{center}
\unitlength\columnwidth
\vspace{-1.5\baselineskip}
\begin{picture}(1,0)
\put(0.21,0){$x$}
\put(0.66,0){$x$}
\put(-0.005,0.23){$y$}
\put(-0.005,0.67){$y$}
\end{picture}
\caption{Maps of normalized error $q_\mathrm{err}/{q_\mathrm{err}^{\max}}\in[0.001,1]$ at the highest resolution $n_x\times n_y=1441 \times 1441$. Left: velocity error $|u|_\mathrm{err}$. Right: pressure error  $p_\mathrm{err}$. Top: forcing without velocity interpolation near the cylinder surface. Bottom: forcing with velocity second-order interpolation at mesh nodes next to the immersed boundary (see Fadlun \textit{et al.} \cite{fadlunetal00}).}
\label{mapmacart}
\end{figure}

Maps of the error field are shown in figure \ref{mapmacart} for the two different forcing techniques. 

For the forcing without interpolation (figure \ref{mapmacart}-top), the dominant errors are located near the cylinder, in its upstream part, up to the separation points. Its leads further downstream to a contamination of the flow in a major part of the wake region. This location of the maximum errors near the separation points on the cylinder is confirmed by the error map of the pressure, with an impact on the pressure solution even far from the body both in the upstream and downstream regions. When an interpolation is used in the forcing (figure \ref{mapmacart}-bottom), the error pattern is drastically modified with a strong error concentration in the near-cylinder region. This more local character of the error is even recovered for the pressure despite the non-local character of this physical quantity. The strongly dominant errors very close to the immersed boundary surface suggest that an improvement of the forcing method, especially for the pressure treatment, could increase significantly the overall accuracy of the solution. The current forcing based on interpolation, even if it is much more accurate than a staircase forcing, still leads to an unbalanced approximation where the immersed boundary treatment contaminates the solution regardless the quality of the discretization far from the body surface.

An interesting remark is that the use of the second-order IBM does not lead to a dramatic increase of the error despite the lack of any fitting between the mesh and the cylinder. For instance, the highest resolutions $n_r\times n_\theta=721 \times 2880$ and $n_x\times n_y=1441\times 1441$ both yield to an error with an order of magnitude of $10^{-4}$ for the velocity. This similar error level is obtained using the same number of degrees of freedom but through a drastic increase of the computational cost (about one order of magnitude) for the body-fitting approach due to the time step limitation associated with the use of a very refined grid in the azimuthal direction. Then, for a given accuracy, the use of an IBM combined with a Cartesian grid is found to be much more computationally efficient compared with the use of a cylindrical grid. However, a significant difference between body fitted and IBM solutions concerns the pressure field that is found to be less accurate in the latter case. Once again, the full access to an accurate reference solution exhibits easily this striking difference. Here, it suggests that an improvement of the immersed boundary modelling could be obtained through a pressure treatment combining accurately the incompressibility condition with the forcing to ensure more realistically no-slip conditions at the cylinder surface.

\section{Conclusion\label{conclusion}}

In this study, an accurate solution of the flow over a circular cylinder at $Re=40$ is provided using a fully pseudo-spectral code. To mimic an infinite computational domain, a new set of boundary conditions is proposed and validated. The numerical solution is obtained up to 50 diameters far from the cylinder. The solution can be known everywhere inside the corresponding computational domain using an interpolation with spectral accuracy. To be user-friendly, the solution is provided as a script for GNU Octave or Matlab software that gives the expected values of velocity components or pressure in any couple of coordinates $(x,y)$ inside the computational domain of the reference simulation.

It is shown how the knowledge of this reference solution can be useful for code validation or development. First, it can be used to define properly the boundary conditions of the problem so that uncertainty associated with the use of open boundary conditions can be eliminated. Then, thanks to its high accuracy, the numerical solution can be assimilated as an exact solution to analyse the convergence of the method or to identify its prevailing errors. The potential benefit of this reference solution is illustrated using two typical codes, one based on a body-fitted grid and another one using an IBM on a Cartesian grid. For the first time in our knowledge, the spatial distribution of errors is exhibited, suggesting a careful consideration of the separation phenomenon, especially when the body is modelled using an IBM. A clear advantage of the reference solution is that it enables comparisons between numerical codes through a rigorous estimation of their actual accuracy. It should be stressed again that this estimation requires the use of the reference solution to provide well defined boundary conditions while avoiding the uncertainty associated with the modelling of an infinite domain.

Our hope is that this very flexible numerical solution will be useful for a wide scientific community in order to inspire or validate new developments, especially in the context of IBM where the second-order accuracy must be overtaken to widen its applications in the field of high-order methods. 

We are aware that the flow configuration is rather simple. For further development, the dipole-wall collision seems to be a good candidate for a non-stationary reference solution that is reachable with the spectral accuracy \cite{keetelsetal07}. On the one hand the vortical sheet layers generated during the collision process are numerically challenging to capture at high Reynolds number, on the other hand their deterministic nature should enable a reliable spatial and temporal convergence study. In the same spirit, the problem of the oblique collision of a ring-vortex with a wall \cite{verziccoandorlandi94} could be useful to assess a numerical code regarding its ability to describe accurately the three-dimensional near-wall dynamics.

\section*{Appendix A. Access to the reference solution\label{database}}

This section describes the method for interpolating the value of the reference solution at any grid point while preserving spectral accuracy and double precision.

To get values of the reference solution at a given location $(x,y)$, first download the archive file \verb|refsolcyl.tgz| and extract the files and directory. The directory should contain the program interpol.m and the data directory containing the reference solution values.

In a file named \verb|grid.dat|, report the couples of values $(x,y)$ where the reference solution has to be interpolated. Then, execute the software Octave or Matlab and enter the command \verb|interpol|. The solution at each selected locations $(x,y)$ is stored in files named \verb|U.dat|, \verb|V.dat| (velocity components), \verb|P.dat| (pressure) and \verb|vort.dat| (vorticity).

Please note that, because of the lack of physical boundary conditions for the pressure, the pressure has been extrapolated with a spline method on the cylinder and on the exterior boundary. Thus a strictly spectral accuracy cannot be warranty for the pressure.
For points inside the cylinder, the value is arbitrarily zero, while for points exterior to the domain ($i.e.$ $r_\infty>40$), the velocity is arbitrarily set to a uniform streamwise flow.

For example the interpolation on the point $(x=3,~ y=3)$ should give the values:\\
$\mathrm{U}=1.0670111347099145e+00$\\
$\mathrm{V}=1.5134054077428072e-02$\\
$\mathrm{P}=-7.0794830815655407e-02$\\
$\mathrm{vort}=-6.5845796777125321e-06$
\\
Small variations starting from to the fourteenth decimal can be observed, depending on the computer used.

The data files and the interpolation program are available on the journal website, also on demand to the corresponding author.



%% References

%\newpage
\bibliographystyle{elsarticle-num}
\bibliography{CF2012,biblio}

\begin{thebibliography}{10}
\expandafter\ifx\csname url\endcsname\relax
  \def\url#1{\texttt{#1}}\fi
\expandafter\ifx\csname urlprefix\endcsname\relax\def\urlprefix{URL }\fi
\expandafter\ifx\csname href\endcsname\relax
  \def\href#1#2{#2} \def\path#1{#1}\fi

\bibitem{tritton59}
D.~J. Tritton, Experiments on the flow past a circular cylinder at low reynolds
  numbers, J. Fluid Mech. 6 (1959) 547--567.

\bibitem{dennis&chang70}
S.~Dennis, G.~Chang, Numerical simulations for steady flow past a circular
  cylinder at {R}eynolds number up to 100, J. Fluid Mech. 42 (1970) 471--489.

\bibitem{coutanceau&bouard77}
M.~Coutanceau, R.~Bouard, Experimental determination of the main features of
  the viscous flow in the wake of a circular cylinder in uniform translation.
  {P}art 1. {S}teady flow., J. Fluid Mech. 79 (1977) 231--256.

\bibitem{fornberg80}
B.~Fornberg, A numerical study of steady viscous flow past a circular cylinder,
  J. Fluid Mech. 98 (1980) 819--855.

\bibitem{he&doolen97}
X.~He, G.~Doolen, Lattice boltzmann method on curvilinear coordinates system:
  Flow around a circular cylinder, J. Comp. Phys. 134 (1997) 306--315.

\bibitem{yeetal99}
T.~Ye, R.~Mittal, H.~Udaykumar, W.~Shyy, An accurate {C}artesian grid method
  for viscous incompressible flows with complex immersed boundaries, J. Comp.
  Phys. 156 (2006) 209--240.

\bibitem{calhoun02}
D.~{\rm Calhoun}, A cartesian grid method for solving the two-dimensional
  streamfunction-vorticity equations in irregular regions, J. Comp. Phys. {\rm
  176} (2002) 231--275.

\bibitem{russel&wang03}
D.~Russel, Z.~Wang, A cartesiand grid method for modeling multiple moving
  objects in {2D} incompressible viscous flow, J. Comp. Phys. 191 (2003)
  177--205.

\bibitem{tseng&ferziger03b}
Y.-H. Tseng, J.~Ferziger, A ghost-cell immersed boundary method for flow in
  complex geometry, J. Comp. Phys. 192 (2007) 593--623.

\bibitem{linnick&fasel05}
M.~Linnick, H.~Fasel, A high-order immersed interface method for simulating
  unsteady incompressible flows on irregular domains, J. Comp. Phys. 204 (2005)
  157--192.

\bibitem{chung06}
M.~Chung, Cartesian cut cell approach for simulating incompressible flows with
  rigid bodies of arbitrary shape, Computers and Fluids 35 (2006) 607--623.

\bibitem{leetal06}
D.~Le, B.~Khoo, J.~Peraire, An immersed interface method for viscous
  incompressible flows involving rigid and flexible boundaries, J. Comp. Phys.
  220 (2006) 109--138.

\bibitem{dingetal07}
H.~Ding, C.~Shu, Q.~Cai, Applications of stencil-adaptive finite difference
  method to incompressible viscous flows with curved boundary, Computers and
  Fluids 36 (2007) 786--793.

\bibitem{taira&colonius07}
K.~Taira, T.~Colonius, The immersed boundary method: A projection approach, J.
  Comp. Phys. 225 (2007) 2118--2137.

\bibitem{Posdziec}
O.~Posdziech, R.~Grundmann, A systematic approach to the numerical calculation
  of fundamental quantities of the two-dimensional flow over a circular
  cylinder, J. of Fluids and Structures 23 (2007) 479--499.

\bibitem{patil&lakshisha09}
D.~Patil, K.~Lakshmisha, Finite volume tvd formulation of lattice boltzmann
  simulation on unstructured mesh, J. Comp. Phys. 228 (2009) 5262--5279.

\bibitem{bouchonetal12}
F.~Bouchon, T.~Dubois, N.~James, A second-order cut-cell method for the
  numerical simulation of {2D} flows past obstacles, Computers and Fluids 65
  (2012) 80--91.

\bibitem{mittal&iaccarino05}
R.~Mittal, G.~Iaccarino, Immersed bondary methods, Ann. Rev. Fluid Mech. {\rm
  37} (2005) 239--261.

\bibitem{chorin68}
A.~Chorin, Numerical simulation of the navier-stokes equations, Math. Comput.
  22 (1968) 745--762.

\bibitem{temam69}
R.~Temam, Sur l'approximation de la solution des \'equations de navier-stokes
  par la m\'ethode des pas fractionnaires {II}., Archiv. Rat. Mech. Anal. 32
  (1969) 377--385.

\bibitem{Botella}
O.~Botella, On the solution of the {N}avier-{S}tokes equations using
  {C}hebyshev projection schemes with third-order accuracy in time, Computers
  \& Fluids 26 (1997) 107--116.

\bibitem{pouxetal11}
A.~Poux, S.~Glockner, M.~Aza\"{i}ez, Improvements on open and traction boundary
  conditions for {N}avier-{S}tokes time-splitting method, J. Comp. Phys.
  230~(10) (2011) 4011--4027.

\bibitem{Pradeep}
D.~S. Pradeep, F.~Hussain, Effects of boundary condition in numerical
  simulations of vortex dynamics, J. Fluid Mech. 516 (2004) 115--124.

\bibitem{Hasan}
N.~Hasan, S.~Anwer, S.~Sanghi, On the outflow boundary condition for external
  incompressible flows: {A} new approach, J. Comp. Phys. 206 (2005) 661--683.

\bibitem{Schlichting}
Schlichting, Boundary layer theory, McGraw-Hill, 1955.

\bibitem{harlow&welch65}
F.~H. Harlow, J.~E. Welch, Numerical calculation of time-dependent viscous
  incompressible flow of fluid with free surface, POF 8~(12) (1965) 2182--2189.

\bibitem{orlandi00}
P.~Orlandi, Fluid Flow Phenomena: A Numerical Toolkit, Springer, 2000.

\bibitem{fadlunetal00}
E.~A. {\rm Fadlun}, R.~{\rm Verzicco}, P.~{\rm Orlandi}, J.~{\rm Mohd-Yusof},
  Combined immersed-boundary finite-difference methods for three-dimensional
  complex flow simulations, J. Comp. Phys. {\rm 161} (2000) 35--60.

\bibitem{keetelsetal07}
G.~Keetels, U.~D'Ortona, W.~Kramer, H.~J.~H. Clercx, K.~Schneider, G.~J.~F. van
  Heijst, Fourier spectral and wavelet solvers for the incompressible
  {N}avier-{S}tokes equations with volume-penalization: Convergence of a
  dipole-wall collision, J. Comp. Phys. \rm 227 (2007) 919--945.

\bibitem{verziccoandorlandi94}
R.~Verzicco, P.~Orlandi, Normal and oblique collisions of a vortex ring with a
  wall, Meccanica 29 (1994) 383--391.

\end{thebibliography}

\end{document}